\documentclass{article}

\usepackage{spconf}

\usepackage[utf8]{inputenc} 
\usepackage[T1]{fontenc}    
\usepackage{hyperref}       
\usepackage{url}            
\usepackage{booktabs}       
\usepackage{amsfonts}       
\usepackage{nicefrac}       
\usepackage{microtype}      
\usepackage{lipsum}
\usepackage{graphicx}
\usepackage{amsmath}
\usepackage{subcaption}
\usepackage{caption}
\usepackage{placeins}
\usepackage{colortbl}
\usepackage{color}
\usepackage{tcolorbox}
\usepackage{multirow}
\graphicspath{ {./figures/} }

\usepackage{bm} 

\ninept
\hypersetup{
    colorlinks = true,
    linkcolor = black,
    anchorcolor = black,
    citecolor = black,
    filecolor = black,
    urlcolor = blue
    }

\title{Collaborative Watermarking for Adversarial Speech Synthesis}

\twoauthors
{Lauri Juvela \sthanks{We acknowledge the computational resources provided by the Aalto Science-IT project.
}  
}{Aalto University, Espoo, Finland \\
\texttt{lauri.juvela@aalto.fi} 
}
{Xin Wang \sthanks{This work was supported by JST CREST Grant JPMJCR18A6 and PRESTO Grant JPMJPR23P9, and MEXT KAKENHI Grant 21H04906.}
}{National Institute of Informatics, Tokyo, Japan \\
\texttt{wangxin@nii.ac.jp}
}

\begin{document}

\onecolumn
{\noindent\Large \textbf{IEEE Copyright Notice}}

${}$

{\noindent\large \copyright 2024 IEEE. 
Personal use of this material is permitted. Permission from IEEE must be obtained for all other uses, in any current or future media, including reprinting/republishing this material for advertising or promotional purposes, creating new collective works, for resale or redistribution to servers or lists, or reuse of any copyrighted component of this work in other works.

${}$

\noindent
This work has been accepted to the IEEE International Conference on Acoustics, Speech and Signal Processing.

\vspace*{\fill}

\noindent Please cite the published version of the paper.
\begin{verbatim}
@inproceedings{juvela24-collaborative-watermarking,
author = {Lauri Juvela and Xin Wang},
title = {Collaborative Watermarking for Adversarial Speech Synthesis},
booktitle = {Proc. ICASSP},
year = {2024},
note = {arXiv preprint arXiv:2309.15224}
}
\end{verbatim}

}
\twocolumn


\maketitle
\begin{abstract}
Advances in neural speech synthesis have brought us technology that is not only close to human naturalness, but is also capable of instant voice cloning with little data, and is  highly accessible with pre-trained models available.
Naturally, the potential flood of generated content raises the need for synthetic speech detection and watermarking. 
Recently, considerable research effort in synthetic speech detection has been related to the Automatic Speaker Verification and Spoofing Countermeasure Challenge (ASVspoof), which focuses on passive countermeasures.
This paper takes a complementary view to generated speech detection: a synthesis system should make an active effort to watermark the generated speech in a way that aids detection by another machine, but remains transparent to a human listener. 
We propose a collaborative training scheme for synthetic speech watermarking and show that a HiFi-GAN neural vocoder collaborating with the ASVspoof 2021 baseline countermeasure models consistently improves detection performance over conventional classifier training. Furthermore, we demonstrate how collaborative training can be paired with augmentation strategies for added robustness against noise and time-stretching. Finally, listening tests demonstrate that collaborative training has little adverse effect on perceptual quality of vocoded speech.



\end{abstract}

\section{Introduction}
\label{sec:intro}

Modern speech synthesis systems have achieved nearly human level naturalness and are capable of zero-shot voice cloning from a few seconds of adaptation data \cite{jia2018-transfer-learning-from-speaker-verification}. 
Open-source implementations, sharing pre-trained models, and good software packaging has made voice cloning with TTS easily accessible also outside the research community \cite{casanova2022yourtts}. Furthermore, the use of voice cloning technology as a service (for a small monthly subscription fee) has recently emerged as a popular past-time on the internet. With this growing user base, the amount of generated speech in the wild is increasing, which poses a risk of casually created misinformation and malicious deepfakes.





Research on audio deepfake detection focuses mostly on passive protection using machine learning methods, which is also referred to as speech anti-spoofing \cite{Wu2015}. This scenario assumes that defenders have no prior knowledge of what attackers will use to generate the audio deepfake. The attackers can use any speech generation models to create the audio deepfake, while the defenders only have a limited number of audio deepfake types in the training set, which does not necessarily cover those from the attackers. Hence, the key question for the defender is how to develop a detection model on the basis of the limited training data and make it generalize to unseen deepfakes. 

Research outcomes from, for example, the ASVspoof \cite{LiuASVspoof2021} and Audio Deep synthesis
Detection (ADD) challenges \cite{yiADD2022a}, have demonstrated some deep learning-based detectors can detect certain unseen deepfakes in the benchmark datasets with error rates smaller than 5\%. However, many detectors were found to be vulnerable to spurious features in the training set \cite{LiuASVspoof2021, muller21_asvspoof, zhang21da_interspeech} and generalize poorly to data from different domains \cite{paul2017generalization, muller2022does}.
Even though the generalization capability may be improved by including more diverse training data, 
the long term equilibrium of this adversarial game of passive audio deepfake detection is yet to be seen.
Attackers can always create audio deepfake from newer and stronger speech generation models.

Generative adversarial networks (GANs) \cite{goodfellow2014-generative-adversarial-nets} take a mirrored perspective on playing the adversarial game of detection and generation. In this setting, a Discriminator network attempts to classify between samples from the real and generated data distributions, while another network, the Generator synthesizes new samples and tries to spoof the discriminator into classifying the generated samples as real. 
The theoretical equilibrium for the GAN game is that the generator learns to match the real data distribution exactly, and discriminating between real and generated samples becomes impossible  \cite{goodfellow2014-generative-adversarial-nets}. Despite practical challenges with finite model capacity, numerics, and unstable training dynamics, GAN-based synthesis methods \cite{Juvela2019-gelp, Yamamoto2019-parallel-wavenet-gan} can be used for realistic speech waveform synthesis. In particular, HiFi-GAN \cite{kong2020-hifi-gan} is widely used as a high-quality neural vocoder in text-to-speech synthesis and voice-conversion.



The adversarial perspective remains highly relevant in defence against black-hat attack scenarios. However, 
the intended use of a generative model is often not malicious, and deceptive realism is merely a side-product of a high-quality system. In these use cases, the generative model provided is likely happy to comply to any regulation and make an active effort watermark their generated model outputs as such. The relevance of watermarking is amplified by the common use of pre-trained models and hosted services: if a model has built-in watermarking that is not trivial to remove and does not affect the perceptual quality of the output, most users will not make an effort to remove the watermark.






This paper proposes a collaborative training scheme that tasks a generative model to watermark its output to be more easily detectable by a specific classifier without degrading the perceptual quality.
The experiments use HiFi-GAN \cite{kong2020-hifi-gan} as the generative model and ASVspoof 2021 challenge \cite{LiuASVspoof2021} baseline countermeasure models as watermark detector models.
Additionally, the detection performance under additive noise and time-stretching is shown to improve by differentiable augmentation during training. Results show that collaborative training consistently improves the detection performance compared to the corresponding passively trained countermeasure model.



\section{Related work}
\label{sec:related_work}


Watermarks can be evaluated using multiple characteristics, including robustness, perceptibility, and applications \cite{peticolas2000-watermarking-schemes-evaluation}. Some applications, such as proof-of-ownership or object identification, require a watermark payload with high enough bit count to encode sufficient information, but this paper focuses on an simple zero-bit watermark: watermark is present in synthetic speech and not present in natural speech.
Sometimes researchers use the terms \textit{watermarking} and \textit{fingerprinting} interchangeably. We use fingerprint to denote the summary of the perceptible information in the carrier speech signal, such as linguistic content, expression, speaker identity, or acoustic channel information. In contrast, a watermark aims to convey hidden information and not be perceptible to the human listener.






Audio watermarking usually consists of an embedder and a detector. The embedder embeds a watermark or message, e.g., pseudo-random numbers or a hash, into the input audio, and the detector verifies the message's existence in a watermarked audio. Typical algorithms using DSP include adding pseudo-noise in the temporal or spectral domain (a.k.a spread spectrum watermarking \cite{cox1996secure}), phase coding \cite{bender1996techniques}, and echo hiding \cite{gruhl1996echo}. 
Some recent methods also implement the embedded and detector using DNNs \cite{pavlovic2022-speech-watermarking-dnn, chen2023wavmark}.

A notable approach that combines DSP and statistics is patchwork watermarking \cite{steinebach2004digitale}. When implemented in the spectral domain, this algorithm embeds a single bit $m$ in the spectrum of an audio frame by selecting two sets of frequency bins and changing their amplitude values in opposite ways. Let $s_{k}$ be the spectral amplitude at the $k$-th frequency bin and ${\mathcal{A}}$ and ${\mathcal{B}}$ be the two sets of frequency bins. 
The algorithm can be written as
\begin{equation}
\begin{cases}
s_i \leftarrow \exp(s_i, {1+d}), \quad s_j \leftarrow \exp(s_j, {1-d}),  \quad\text{if } m = 1 \\
s_i \leftarrow \exp(s_i, {1-d}), \quad s_j \leftarrow \exp(s_j,{1+d}),   \quad\text{if } m = 0
\end{cases},
\end{equation}
where ${i\in\mathcal{A}}$, ${j\in\mathcal{B}}$, and $d$ is a strength parameter. 
The general idea to detect the single bit of message is to compare the mean values of the two sets 
\begin{equation}
\begin{cases}
\widehat{m} = 1, \text{if } \frac{1}{|\mathcal{A}|}\sum_{i\in\mathcal{A}}s_i -\frac{1}{|\mathcal{B}|}\sum_{j\in\mathcal{B}}s_j > 0
\\
\widehat{m} = 0, \text{else} 
\end{cases}.
\end{equation}
A larger strength $d$ value increases the robustness against distortions but degrades the quality of the watermarked audio.

For generative models in the image domain, watermarking the training data with standard DSP techniques has been shown to transfer to GAN outputs
\cite{Yu2021_ICCV-fingerprinting-generative-models-rooting-attribution-in-training-data}.
%
Yu \textit{et al.} \cite{yu2022-responsible-disclosure-fingerprinting} propose a jointly trained watermark encoder-decoder scheme for GANs. This is closely related to our work, both in motivation (``from \textit{passive classifiers} to \textit{proactive fingerprinting}''), and implementation (their Generator acts as a watermark embedding model). Main differences are different domains and our work adds augmentation for robustness.

\section{Proposed method}
\label{sec:proposed-method}

\begin{figure}
    \centering
    \includegraphics[width=0.9\linewidth]{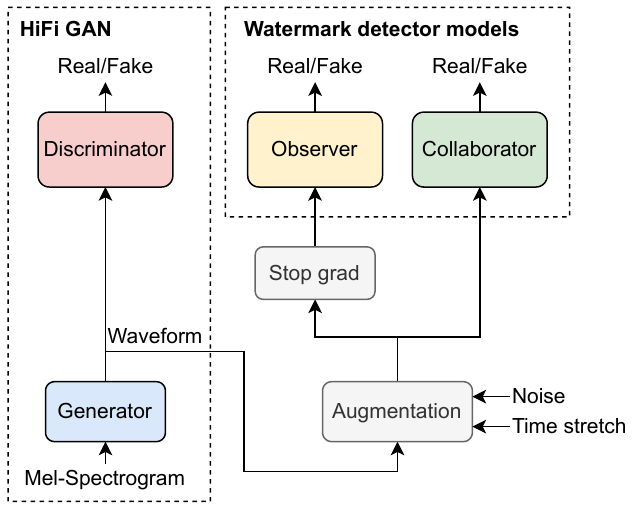}
    \caption{A Generator can view detector models in three distinct roles: fool the Discriminator to produce more realistic samples, ignore the Observer, or help the Collaborator to extract a watermark from generated speech.   
    }
    \label{fig:system-diagram}
\end{figure}

Figure \ref{fig:system-diagram} depicts an overview of the system. A Generator model takes a mel-spectrogram as input and outputs a corresponding synthetic speech waveform. Detector models all try to classify between real and generated speech, and the training dynamics change based on which role the classifier takes:
\begin{enumerate}
    \item \textbf{Discriminator} is adversarial to the Generator. The Generator attempts to fool the discriminator into classifying generated samples as real.
    \item \textbf{Observer} acts as a passive detector. Gradient flow from the Observer to Generator is detached. This corresponds to traditional ASVspoof countermeasure training.
    \item \textbf{Collaborator} shares its objective with the Generator. The pair attempt to embed a watermark into the generated signal in a way that aids the classifier, but does not hinder Generator's other objectives.
\end{enumerate}

%

%
Notably, the general approach is not specific to GANs or ASVspoof detector models, and can be adapted to a wide range of detectors and generative models. 
For simplicity, we chose to limit the experiments to neural vocoding instead of end-to-end TTS. Previous research has demonstrated that training ASV Spoofing countermeasures on vocoded speech transfers well to TTS systems using the same vocoders
\cite{wang2023-spoofing-countermeasure-neural-vocoder}.

\subsection{Loss functions}

Generator and Discriminator loss functions and architectures follow the HiFi-GAN recipe \cite{kong2020-hifi-gan}. We use the large Generator model configuration (V1) and base the experiments on the official implementation\footnote{\scriptsize\url{https://github.com/jik876/hifi-gan}}.

Denote generated speech signal as $x_\text{gen} = G(m)$, where $G$ is the Generator model, and $m$ is the mel-spectrum of the target waveform.
Discriminator, $D$, is trained with the least-squares GAN loss function, with the target score at one for real samples and at zero for generated
\begin{equation}
    \mathcal{L}_{D} = \mathbb{E} \left[(D(x_\text{real})- 1)^2 + D(x_\text{gen})^2 \right], 
\end{equation}
where the expectation, $\mathbb{E}$, is approximated by minibatch averages over batch elements and timesteps. Each of the sub-discriminators in the HiFi-GAN discriminator ensemble share the same objective. 
Similarly, the Generator loss functions are from HiFi-GAN, comprising the adversarial loss
\begin{equation}
    \mathcal{L}_\text{G-adv} = \mathbb{E} \left[ \left( D(x_\text{gen} ) - 1 \right)^2 \right], 
\end{equation}
feature matching loss at each hidden activation $D_{i}(\cdot)$ 
\begin{equation}
    \mathcal{L}_\text{G-FM} =  \sum_{i} \mathbb{E} \left[ \left(D_{i}(x_\text{real})- D_{i}(x_\text{gen}) \right)^2 \right], 
\end{equation}
and L1 regression loss on log-mel-spectrograms
\begin{equation}
    \mathcal{L}_\text{G-mel} = \mathbb{E} \left[ |  \log (\bm{M} \cdot | \text{STFT} (x_\text{real}) |) -  \log ( \bm{M} \cdot | \text{STFT} (x_\text{gen}) |) \ \right], 
\end{equation}
where $\bm{M}$ is a tensorized mel-filterbank matrix. 

\newcommand\WM{\mathit{WM}}

The watermark detector model, $\WM$, has exactly the same objective as $D$: assign a high score to $x_\text{real}$ and a low score to $x_\text{gen}$
\begin{equation}
    \mathcal{L}_{\WM} = \mathbb{E} \left[(\WM(x_\text{real} )- 1)^2 + \WM(x_\text{gen})^2 \right]. 
\end{equation}
$G$ can either share the $\WM$ objective or ignore it. These two scenarios are called Collaborator and Observer, respectively. 

In practice, the system is trained by alternating minibatches that switch between the $D$ objectives and the joint objective of $G$ and $\WM$. In Observer mode, the gradient flow from $\WM$ to $G$ is detached and $\mathcal{L}_{\WM}$ has no effect on $G$.


\subsection{Watermark detector models}


Both ASVspoof 2021 baseline models are designed to operate at at 16\,kHz sample rate, whereas the HiFi-GAN defaults to  22.05\,kHz sample rate. To maintain compatibility with pre-trained models, we use a differentiable resampling layer between the Generator and watermark detector models. The resampler uses a Hann-windowed sinc interpolation filter with six filter taps. 

Pre-trained models use batch normalization in evaluation mode with stored statistics. For joint training, we removed batch normalization layers after identifying they led to poor generalization with collaborative training. Recent research has found that removing batch normalization boosts adversarial training \cite{wang2022-removing-batch-norm-boosts-adversarial-training}, and we speculate the underlying cause may be related due to similar mathematics of adversarial and collaborative training.






\textbf{LFCC-LCNN} is the first detector used in this study, as the baseline of the ASVspoof 2021 challenge\footnote{\scriptsize\url{https://github.com/asvspoof-challenge/2021}}. Its front end extracts linear frequency cepstrum coefficients (LFCC) \cite{sahidullah2015comparison}, which is similar to the Mel frequency cepstrum coefficients but uses filters placed in equal sizes on a linear scale. The frame length and hop size are 20 and 10\,ms, respectively. The FFT size is 1,024, and the number of dimensions is 60, including delta and delta-delta components. The maximum frequency covered by the filter is half of the Nyquist frequency. 
The back end is a light convolution neural network (LCNN) \cite{wuLightCNNDeep2018}, followed by two recurrent layers using long-short term memory units, global average pooling, and a linear output layer \cite{wangComparativeStudyRecent2021}.


\textbf{RawNet2} is another ASVspoof 2021 baseline used in this study is  \cite{tak2020end}. RawNet2 uses convolution (in DNNs) to implement the differentiable band-pass filters. The convolution kernel, or the filter coefficients, are pre-calculated in the same manner as the windowed-sinc filters, and the cut-off frequencies are co-located with the filter bank in MFCC. The input waveform is processed by the convolution layer and transformed by multiple blocks with 2D convolution, max pooling, and filter-wise feature map scaling. The hidden features are then processed by a recurrent layer with gated recurrent units and a linear output layer. Its configuration follows the official implementation, except that the batch normalization layer is removed. The input waveform is padded or truncated to a fixed length of 65,536.

\subsection{Differentiable augmentation} 

We apply two differentiable augmentation techniques to improve the robustness of the proposed collaborative watermarking method.
First, time-stretching is implemented using linear interpolation. The time-scale factor is randomized uniformly between time-scale factors 0.9 and 1.1 and kept constant for each mini-batch.
Second, additive noise samples are randomly drawn from the MUSAN database \cite{musan2015} noise subset. For simplicity, we keep the noise level constant at 10\,dB SNR.
More fine-grained analysis on the effect of varying noise levels is left as future work.

\begin{table*}[thb]
    \centering
   \caption{EER (\%) of experiment systems in clean, stretch, noise, and stretch plus noise (S+N) testing conditions. A darker cell color indicates a higher EER value. Each EER was averaged over 20 independent rounds of generation and evaluation. Error rates (\%) of spectral-domain patchwork watermarking are shown at the top for reference.}
   \begin{tabular}{rrrrrrrrrrr}
\toprule
&   &  &      Clean      &     Stretch     &      Noise  & S+N    \\
\cmidrule(lr){4-7}
\multicolumn{3}{r}{Spectral patchwork watermarking} & 0.21 & 32.65 & 85.29 & 98.25 \\
\midrule
 \multicolumn{3}{c}{WM configuration} & \multicolumn{4}{c}{LFCC-LCNN} & \multicolumn{4}{c}{RawNet} \\
 \cmidrule(lr){4-7}\cmidrule(lr){8-11}
Training & Augmentation  & Role                                          
&      Clean      &     Stretch     &      Noise      & S + N &      Clean      &     Stretch     &      Noise      & S + N\\ 
\midrule
\multirow{4}{*}{Joint} & \multirow{2}{*}{None}  
        & collaborator          & \cellcolor[rgb]{0.94, 0.94, 0.94} 1.61 & \cellcolor[rgb]{0.92, 0.92, 0.92} 2.95 & \cellcolor[rgb]{0.78, 0.78, 0.78} 34.66 & \cellcolor[rgb]{0.76, 0.76, 0.76} 42.90 & \cellcolor[rgb]{0.98, 0.98, 0.98} 0.12 & \cellcolor[rgb]{0.76, 0.76, 0.76} 43.25 & \cellcolor[rgb]{0.76, 0.76, 0.76} 44.33 & \cellcolor[rgb]{0.75, 0.75, 0.75} 48.62\\ 
&  & observer            
& \cellcolor[rgb]{0.92, 0.92, 0.92} 3.46 & \cellcolor[rgb]{0.90, 0.90, 0.90} 5.13 & \cellcolor[rgb]{0.77, 0.77, 0.77} 37.87 & \cellcolor[rgb]{0.75, 0.75, 0.75} 46.20 & \cellcolor[rgb]{0.97, 0.97, 0.97} 0.34 & \cellcolor[rgb]{0.83, 0.83, 0.83} 17.86 & \cellcolor[rgb]{0.78, 0.78, 0.78} 35.35 & \cellcolor[rgb]{0.75, 0.75, 0.75} 49.65\\ 
\cmidrule{2-11}
& \multirow{2}{*}{Stretch + noise} & collaborator    
& \cellcolor[rgb]{0.95, 0.95, 0.95} 1.05 & \cellcolor[rgb]{0.94, 0.94, 0.94} 1.38 & \cellcolor[rgb]{0.85, 0.85, 0.85} 15.25 & \cellcolor[rgb]{0.78, 0.78, 0.78} 34.94 & \cellcolor[rgb]{0.95, 0.95, 0.95} 1.36 & \cellcolor[rgb]{0.93, 0.93, 0.93} 2.33 & \cellcolor[rgb]{0.91, 0.91, 0.91} 4.03 & \cellcolor[rgb]{0.74, 0.74, 0.74} 54.10\\ 
& & observer
& \cellcolor[rgb]{0.92, 0.92, 0.92} 3.73 & \cellcolor[rgb]{0.91, 0.91, 0.91} 4.24 & \cellcolor[rgb]{0.80, 0.80, 0.80} 27.72 & \cellcolor[rgb]{0.76, 0.76, 0.76} 41.35 & \cellcolor[rgb]{0.92, 0.92, 0.92} 3.72 & \cellcolor[rgb]{0.90, 0.90, 0.90} 4.87 & \cellcolor[rgb]{0.85, 0.85, 0.85} 13.31 & \cellcolor[rgb]{0.75, 0.75, 0.75} 46.48\\ 
\midrule
\multirow{3}{*}{Pre-trained} & \multirow{2}{*}{None} & collaborator       
& \cellcolor[rgb]{0.83, 0.83, 0.83} 17.73 & \cellcolor[rgb]{0.82, 0.82, 0.82} 20.87 & \cellcolor[rgb]{0.76, 0.76, 0.76} 42.13 & \cellcolor[rgb]{0.76, 0.76, 0.76} 45.19 & \cellcolor[rgb]{0.87, 0.87, 0.87} 10.64 & \cellcolor[rgb]{0.82, 0.82, 0.82} 21.36 & \cellcolor[rgb]{0.80, 0.80, 0.80} 28.32 & \cellcolor[rgb]{0.74, 0.74, 0.74} 51.48\\
& & observer         & \cellcolor[rgb]{0.75, 0.75, 0.75} 49.47 & \cellcolor[rgb]{0.75, 0.75, 0.75} 49.50 & \cellcolor[rgb]{0.75, 0.75, 0.75} 49.05 & \cellcolor[rgb]{0.75, 0.75, 0.75} 49.56 & \cellcolor[rgb]{0.75, 0.75, 0.75} 47.76 & \cellcolor[rgb]{0.75, 0.75, 0.75} 48.35 & \cellcolor[rgb]{0.75, 0.75, 0.75} 48.55 & \cellcolor[rgb]{0.75, 0.75, 0.75} 50.15\\ 
\cmidrule{2-11}
& Stretch + noise & collaborator 
& \cellcolor[rgb]{0.78, 0.78, 0.78} 32.83 & \cellcolor[rgb]{0.78, 0.78, 0.78} 32.79 & \cellcolor[rgb]{0.77, 0.77, 0.77} 40.79 & \cellcolor[rgb]{0.76, 0.76, 0.76} 44.37 & \cellcolor[rgb]{0.75, 0.75, 0.75} 45.91 & \cellcolor[rgb]{0.75, 0.75, 0.75} 46.75 & \cellcolor[rgb]{0.75, 0.75, 0.75} 47.18 & \cellcolor[rgb]{0.75, 0.75, 0.75} 48.40\\ 
\bottomrule
\end{tabular}

\label{tab:eer_all}
\end{table*}
\section{Experiments}

\subsection{Datasets}

The Voice Cloning Toolkit (VCTK) \cite{Yamagishi2019-VCTK} corpus is used for all the speech data in the experiments.
VCTK has a data split protocol for ASVspoof aimed at evaluation purposes \cite{LiuASVspoof2021}, but we deemed the training set too small for training the synthesis system. Instead, we opted for a custom
80-10-10\% split to training, validation, and test sets. Split details are provided with the source code.
Each subset has distinct speakers, but has overlaps in text content. 
Noise data for augmentation consists of the noise subset from the MUSAN database \cite{musan2015}.  We apply a 80-10-10\% split and use unseen noise samples during testing.

\subsection{Training details}

Following HiFi-GAN, we use the AdamW optimizer with learning rate 2e-4, $\beta_1$ = 0.8, $\beta_2$ = 0.99 and learning rate decay with exponential decay factor 0.999. Training segements are padded or randomly cropped to 8,192 samples with LFCC-LCNN and 65,536 samples with RawNet.
In Collaborator models, the gradients are allowed to flow back from the $\WM$ to $G$, while in Observer models the gradient is detached.
%
Each model configuration is trained for 50 epochs, amounting to approximately 110k iterations. Note that the HiFi-GAN paper \cite{kong2020-hifi-gan} reports training for 2.5M iterations, which does yield higher synthesis quality. This work focuses on demonstrating the relative benefit of collaborative training over baseline observer training in many scenarios.
We believe the current setup offers a reasonable trade-off between the computational cost of training many systems, and the perceptual quality of generated speech.  
The various model configurations were trained on available hardware from a pool of GPUs consisting of NVIDIA A100, V100 and P100 models.

\subsection{Patchwork watermark baseline}

As a DSP baseline, patchwork watermarking is applied to vocoded speech from a HiFi-GAN trained without detector collaboration.
The patchwork watermarking algorithm described in Section~\ref{sec:related_work} is included for reference. The open-sourced toolkit Audiowmark \footnote{\scriptsize\url{https://github.com/swesterfeld/audiowmark}} is used due to its high-quality implementation with error-correction in watermarking detection. A fixed-length 128 bit text string is watermarked. The hyper-parameter $d$ was selected for each test utterance through grid search. Given the range between 0.01 and 0.2, the value of $d$ was decided so that the detector outputs the correct watermark without other hypothesis. During test conditions, the watermarked audio is added with noise or stretched before sent to the watermark detector.

\subsection{Listening test}
\label{sec:listening-test}

\begin{table}[ht!]
    \centering
    \footnotesize
    \caption{Subjective MOS results with confidence interval (95\%). Proposed models use WM as collaborator.}
    \begin{tabular}{lrrrr}
    \toprule
    \multirow{3}{*}{\rotatebox{0}{Ref.}} & \multicolumn{2}{r}{Natural recording} & $4.13 \pm 0.11$ \\
    & \multicolumn{2}{r}{Spectral patchwork} & $3.49 \pm 0.13$ \\
    & \multicolumn{2}{r}{Baseline HiFi-GAN} & $3.54 \pm 0.13$ \\ 
    \midrule
    \multirow{6}{*}{\rotatebox{90}{Proposed}} & Training & Augmentation & LFCC-LCNN & RawNet  \\
    \cmidrule{2-5}
    & \multirow{2}{*}{Joint} & None & $3.38 \pm 0.17$ & $3.59 \pm 0.13$ \\
    &  & Stretch + noise  & $3.46 \pm 0.13$ & $3.60 \pm 0.13$ \\
    \cmidrule{2-5}
    & \multirow{2}{*}{Pre-trained} & None  & $3.58 \pm 0.13$ & $3.51 \pm 0.13$ \\
    &  & Stretch + noise  & $3.46 \pm 0.14$ & $3.67 \pm 0.12$ \\
    \bottomrule
    \end{tabular}
    \label{tab:mos}
\end{table}

We conducted a mean opinion score (MOS) test to evaluate the perceptual effect of the proposed method.
Listeners were asked to rate the naturalness of the presented speech samples on a five-point scale ranging from 1 (bad) to 5 (excellent).
The listening test stimuli consist of 50 utterances selected randomly from test set. 
After pooling together test utterances from each system, the stimuli were randomly batched to 100-sample listening sessions, and each batch was rated by at least five listeners on the Prolific crowd-sourcing platform.
%
After screening, 3884 total ratings by 35 listeners were used in the analysis.
Table \ref{tab:mos} shows MOS values with t-statistic based confidence intervals. Differences between vocoded  systems are not statistically significant.



\section{Results and Discussion}

Table \ref{tab:eer_all} displays test set equal error rates (EER) for a range of systems. 
When training watermark detector models jointly with the generator, collaborative training consistently outperforms conventional observer training. This remains consistent over the detector model type, optional augmentation during training, and different test conditions, from clean to time-stretching and/or additive noise.
All LFCC-LCNN configurations struggle with addivive noise, whereas
collaborative training with RawNet and data augmentation still performs well under noise (EER 4.03). However, combining noise with time-stretching remains a challenging for both LFCC-LCNN and RawNet detectors, even when the models were trained with matching data augmentation.

Pre-trained models struggle as watermark detectors, as listed in the bottom part of Table \ref{tab:eer_all}.
In collaborative training, the synthesis model manages to embed some detectable information in clean conditions, but detection performance deteriorates when noise is added.
In a zero-shot scenario (i.e., pre-trainer observer) the performance is near chance level for both LFCC-LCNN and RawNet. It appears that the Generator's attempt to fool the Discriminator transfers to the ASVspoof pre-trained baselines, which have not been trained on HiFi-GAN vocoded speech.

The DSP-based watermarking algorithm achieves a low error rate in the clean condition since the strength $d$ was decided to allow perfect watermark detection in this condition. However, there are a few utterances in which the embedded watermark failed to be detected given the largest $d$. Its error rate increases when the watermarked audio is stretched, even when the provided compensation for playback was used. The performance degrades dramatically when additive noise is applied.
Note that the patchwork baseline is not one-to-one comparable to the detector models, since baseline watermark uses a 128-bit payload and does not output detection scores required for EER calculation.

A central limitation of the current experimental setup is the focus on neural vocoding. Real-world voice cloning applications use a full text-to-speech or voice conversion system. Nevertheless,
we are optimistic on the transferability of training on vocoded speech based on recent results in the ASVspoof scenario
\cite{wang2023-spoofing-countermeasure-neural-vocoder}.
%
%
However, security critical applications still need to follow an adversarial countermeasure protocol.
Collaborative watermarking is only useful when the generative model user has an incentive to watermark their model outputs.
%
Furthermore, collaborative training is not limited to GANs. Any deep generative model with differentiable sampling (including diffusion models \cite{chen2021-wavegrad, kong2021diffwave}) can collaborate with a detector model to improve the odds of detection.
%


\section{Conclusions}

This paper proposes synthetic speech watermarking scheme based on collaborative training between a generative synthesis model and a watermark detector. The results show that collaborative training consistently improves detection performance compared to baseline passive countermeasure training. 
Source code and demonstration samples are available for readers\footnote{\scriptsize\url{https://ljuvela.github.io/CollaborativeWatermarkingDemo/}}.
Future work includes extending the study to full text-to-speech voice cloning systems, and developing  specialized architectures for the speech watermarking task. Further, more informative watermarks are appealing: it would be useful to know who generated the samples, or what data was used to train the model.

\newpage
\FloatBarrier

\bibliographystyle{IEEE} 

\bibliography{references}  

\end{document}